\documentclass{article}

\PassOptionsToPackage{numbers, compress}{natbib}

\usepackage[preprint]{neurips_2026}

\usepackage[utf8]{inputenc} 
\usepackage[T1]{fontenc}    
\usepackage{hyperref}       
\usepackage{url}            
\usepackage{booktabs}       
\usepackage{amsfonts}       
\usepackage{nicefrac}       
\usepackage{microtype}      
\usepackage{xcolor}         

\usepackage{amsmath}
\usepackage{svg}
\usepackage{multirow}
\usepackage{soul}           
\usepackage{wrapfig}        
\usepackage{pifont}

\DeclareMathOperator{\tr}{tr}

\title{\textsc{HyperShape}: Hyperelasticity Across Diverse Shapes}

\author{%
    Leo Widmer\\
    Department of Biomedical Engineering\\
    University of Basel\\
    \texttt{leo.widmer@unibas.ch}\\
    \And
    Sidaty El Hadramy\\
    Department of Biomedical Engineering\\
    University of Basel\\
    \texttt{sidaty.elhadramy@unibas.ch}\\
    \And
    St\'ephane Cotin\\
    MIMESIS Team\\
    Inria, Strasbourg, France\\
    \texttt{stephane.cotin@inria.fr}\\
    \And
    Philippe Claude Cattin\\
    Department of Biomedical Engineering\\
    University of Basel\\
    \texttt{philippe.cattin@unibas.ch}
}

\begin{document}

\maketitle

\begin{figure}[!h]
    \centering
    \includegraphics[width=\linewidth]{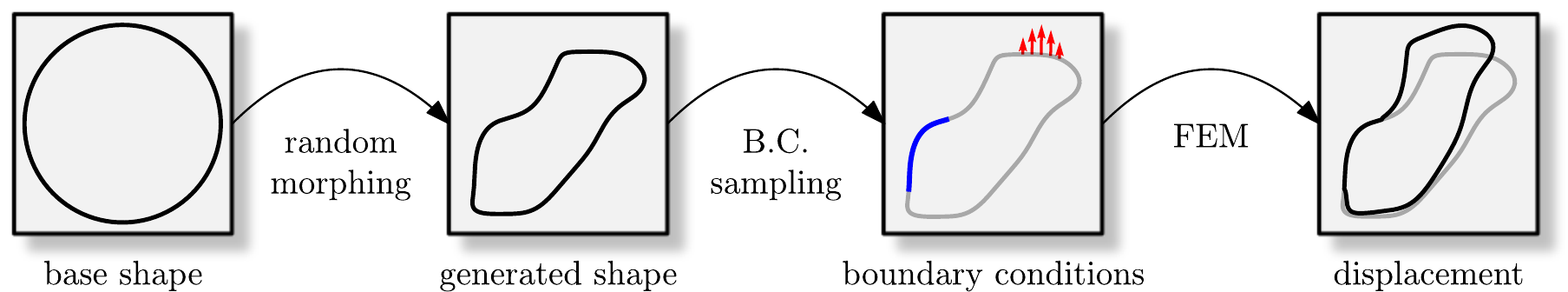}
    \caption{Overview of \textsc{\textbf{HyperShape}} data generation pipeline. A base shape (e.g., circle) is morphed into a synthetic shape of controllable complexity, Dirichlet (\textcolor{blue}{blue}) and Neumann (\textcolor{red}{red}) boundary conditions are randomly assigned, and a FEM simulation is performed under the applied traction force to produce the resulting displacement field.}
    \label{fig:hypershape}
\end{figure}

\begin{abstract}

Hyperelastic deformations are highly sensitive to domain geometry and boundary conditions, making generalization across both a critical capability for neural operators applied to these problems. However, existing benchmarks for neural operators on hyperelasticity rely on simple or few geometries, which makes it difficult to assess this capability rigorously. To address this gap, we introduce \textsc{\textbf{HyperShape}}, an extensible framework designed to generate synthetic shapes and their corresponding hyperelastic simulation data, producing a suite of 2D and 3D datasets with adjustable complexity and controllable shape variations. This design enables systematic assessment of generalization across in-distribution, out-of-distribution, and synthetic-to-real transfer settings. Using this framework, we evaluated the performance of several state-of-the-art neural operators over diverse shape distributions. Our findings reveal that neural operators perform well on simple shapes but struggle as shape complexity, geometric diversity, and boundary condition variability increase, requiring large amounts of training data in such regimes. Performance degrades consistently and predictably with geometric complexity highlighting the need for further model development. As an open and extensible benchmark, \textsc{\textbf{HyperShape}} is designed to grow alongside the field: new geometries, material models, loading conditions, and evaluation settings can be easily incorporated to validate hyperelastic surrogate models.

  
\end{abstract}

\section{Introduction}
\label{sec:intro}

Numerous engineering tasks require accurate modeling of deformable, nonlinear structures subjected to complex loads and boundary conditions, often with an emphasis on real-time or near real-time computation \cite{Polenghi2018-bm}. Such requirements arise across fields such as biomechanics \cite{Holzapfel2025-jp}, materials science \cite{Freutel2014-da}, and interactive system development \cite{Courtecuisse2010-nx}, where both \textbf{precision} and \textbf{computational efficiency} are critical. In this context, hyperelastic simulations are well-established in biomechanics and support a variety of medical applications \cite{Cotin1999-ue}. Hyperelastic models are commonly employed to simulate soft tissues, forming the basis for surgical training simulators \cite{Kojanazarova2025-wq}, preoperative planning \cite{delingette_soft_2004}, intraoperative guidance systems \cite{haouchine_image-guided_2013, suwelack_physics-based_2014, Hadramy2024-pt}, and can be used to solve inverse problems, such as estimating material parameters or identifying boundary conditions \cite{olson_inverse_2019, El_Hadramy2026-dp}. To perform hyperelastic simulations, one must solve the governing partial differential equations (PDEs) over the domain while accounting for the material’s nonlinear constitutive behavior, the applied boundary conditions, and external loads. A variety of numerical solvers have been developed to address this \cite{Matos2023-az, Brunet2019-gk}. Among these, the finite element method (FEM) remains the most widely used due to its ability to model complex geometries and material responses \cite{Freutel2014-da}. However, to meet real-time requirements, these methods often require substantial model simplifications to enable parallelization or reduce the number of degrees of freedom, which can lead to loss of accuracy in the resulting approximate solutions \cite{nguyen_systematic_2020}.


In recent years, deep learning–based approaches have emerged as an alternative to traditional numerical solvers, offering an interesting trade-off between computational speed and accuracy \cite{Herrmann2024-sq}. Among the earliest of these methods are physics-informed neural networks (PINNs) \cite{Raissi2019-xp}, which embed the governing PDEs directly into the loss function. PINNs are trained to solve a single forward or inverse problem associated with a given geometry and boundary conditions. Building on this, patient-specific data-driven approaches were later proposed, in which deep neural networks (CNNs \cite{mendizabal_simulation_2020}, GNNs \cite{deshpande_magnet_2024}) are trained on a predefined geometry to generalize across different loading conditions or boundary settings. This allows them to predict multiple deformation states for the same anatomical shape. Both approaches share the advantage of being fully differentiable, which makes them particularly attractive for inverse problems \cite{olson_inverse_2019, El_Hadramy2026-dp}. However, a key limitation is that they are tied to either a specific problem setup or a fixed geometry; thus, extending them to new cases requires additional training, limiting their general applicability. A further class of methods is data-driven approaches that generalize across local variations within a distribution of geometries \cite{el_hadramy_deform_2025}. These methods capture small geometric variations from data and can only predict deformations for unseen samples drawn from the same distribution. This improves generalization compared to patient-specific models, but performance still degrades on out-of-distribution shapes.

\begin{wrapfigure}{r}{0.3\textwidth}
    \centering
    \includegraphics[width=\linewidth]{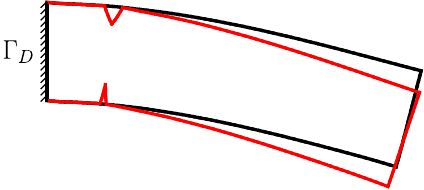}
    \caption{Illustrative example showing the influence of geometry on the solution. Both beams are fixed on the left and pulled with the same force. The only difference is a small local change in the red beam's geometry, which leads to a significantly different global deformation response.}
    \label{fig:beam-with-notch}
\end{wrapfigure}

Many recent works have therefore focused on developing models capable of generalizing across multiple shapes. In this context, neural operators have gained significant attention due to their ability to learn mappings between function spaces, with the promise of generalizing over distributions of PDE solutions
\cite{kovachki_neural_2024, boulle_chapter_2024}. Among the most representative approaches are Fourier Neural Operator (FNO) \cite{li_fourier_2021,kossaifi_multi-grid_2023, duprez_phi-fem-fno_2025}, which leverages spectral convolutions to capture global interactions. The Convolutional Neural Operator (CNO)\cite{raonic_convolutional_2023} extends convolutional architectures to operator learning, while Graph Neural Operators \cite{Li2020-hr} adapt the framework to irregular domains by relying on graph-based representations. In parallel, architectures inspired by image processing, such as U-Net–based neural operators \cite{ronneberger_u-net_2015}, have also been explored. More recent point-cloud-based approaches include  GINO \cite{li_geometry-informed_2023}, Transolver \cite{wu_transolver_2024}, and GAOT \cite{wen_geometry_2026}, which aim to handle complex geometries and unstructured data more flexibly.

\begin{figure}
    \centering
    \includegraphics[width=\linewidth]{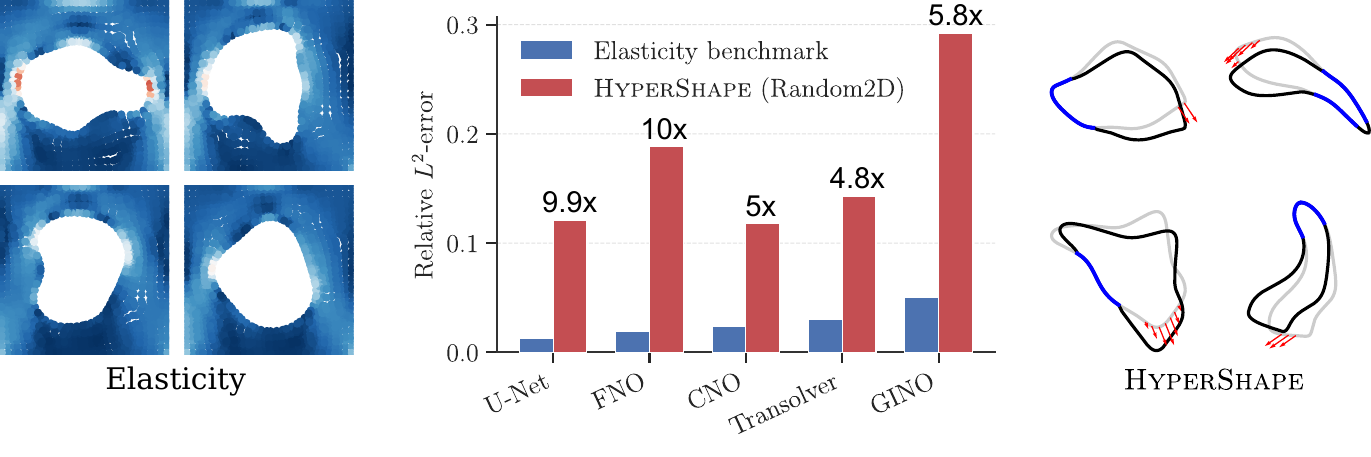}
    \caption{Comparison between the Elasticity benchmark \cite{li_fourier_2024} and \textbf{\textsc{HyperShape}}. In Elasticity (left), all samples share the same geometry (a square with a hole), with fixed Dirichlet boundary conditions (zero displacement at the bottom) and identical traction applied at the top. In contrast, \textbf{\textsc{HyperShape}} (right) features varying geometries, boundary conditions, and forces, making the problem more realistic and challenging. This is reflected in the histograms (middle): while state-of-the-art neural operator models perform well on Elasticity (in \textcolor{blue}{blue}), their errors increase significantly on the more diverse \textbf{Random2D} dataset generated with \textbf{\textsc{HyperShape}} (in \textcolor{red}{red}).}
    \label{fig:elasticity-benchmark-comparison}
\end{figure}

While neural operators offer strong advantages in learning solution mappings across families of PDEs, their application to hyperelastic simulations remains relatively limited. Most existing studies focus on standard benchmark problems commonly used in the literature, such as Darcy flow, Burgers’ equation, or the incompressible Navier–Stokes equations\cite{li_fourier_2021}, which use simple geometries and controlled settings. In contrast, hyperelasticity exhibits a large \textbf{sensitivity to geometric variations}: even small, localized changes in the domain can induce significant differences in the global deformation response, as illustrated in Figure~\ref{fig:beam-with-notch}. Although many neural operator frameworks claim the ability to generalize across geometries \cite{li_geometry-informed_2023, kovachki_neural_2024, wen_geometry_2026}, this assumption is particularly difficult to sustain in the context of hyperelasticity. Current benchmarks include limited geometric variability and therefore fail to capture the strong sensitivity of hyperelastic responses to shape changes. One of the most commonly used hyperelasticity benchmarks \cite{li_fourier_2024} uses a distribution of shapes consisting of squares with randomly shaped holes and the \textbf{same boundary conditions for every geometry} (applied to the top and bottom of the square). The left-most image in Figure~\ref{fig:elasticity-benchmark-comparison} illustrates example geometries from the Elasticity benchmark \cite{li_fourier_2024}, while the blue histograms highlight the strong performance of state-of-the-art models on this dataset. In contrast, the performance of the same models degrades significantly on complex irregular shapes with varying boundary conditions and force loads, as illustrated in the right-most image and reflected by the red histograms. These observations suggest that \textbf{ existing benchmarks may considerably overestimate the robustness and generalization capabilities of current neural operators} by relying on overly simplified and homogeneous settings. This motivates the need for more realistic and controllable benchmarks that capture the variability encountered in practical applications.

This paper introduces \textbf{\textsc{HyperShape}}, a controllable and versatile method for generating synthetic shapes and a suite of 2D and 3D hyperelasticity datasets designed to evaluate neural operators across varying geometries, boundary conditions, and distribution shifts. The shape generation process can be adjusted and tailored to a specific application or to change the difficulty level of the problem for evaluation and failure-mode analysis.  Our contributions are the following.
\begin{itemize}
    \item We introduce \textsc{\textbf{HyperShape}}, a controllable synthetic shape generation framework, along with a suite of 2D and 3D hyperelasticity datasets of adjustable complexity, released with the corresponding generation code.
    \item We evaluated state-of-the-art neural operator architectures on various datasets generated using \textsc{\textbf{HyperShape}}, covering in-distribution, out-of-distribution, and synthetic-to-real transfer settings.
    \item We find that current neural operators struggle to handle complex shape distributions, highlighting the need for further model development.
    \item We propose an analysis of geometry-dependent failure modes, data efficiency, and the conditions under which synthetic pretraining transfers to anatomical geometries.
\end{itemize}

The paper is organized as follows. Section~\ref{sec:prob_formulation} introduces the hyperelastic problem formulation. Section~\ref{sec:hypershape} presents \textsc{\textbf{HyperShape}}, with details about shape sampling, boundary conditions, data representation, evaluation metrics, and the generation of datasets with varying levels of difficulty. Section~\ref{sec:experiments} provides benchmarking of state-of-the-art neural operators on 2D/3D datasets and discuss the main findings. Finally, Section~\ref{sec:conclusion} concludes the paper.

\section{Problem formulation}
\label{sec:prob_formulation}

Let $\Omega$ be a domain representing the geometry of interest, $\Gamma_D\subseteq\partial\Omega$  and $\Gamma_N=\partial\Omega\setminus\Gamma_D$ are respectively the Dirichlet and Neumann boundaries (see Figure~\ref{fig:domain_example}). We aim at solving the boundary value problem by finding a displacement field $u:\Omega\to\mathbb{R}^d$ that satisfies:

\begin{align}\label{eq:bvp}
\begin{cases}
\begin{aligned}
    -\Delta P(u)&=f,&&\text{in }\Omega,\\
    u&=u_D,&&\text{in }\Gamma_D,\\
    P(u)\cdot \vec{n}&=t,&&\text{in }\Gamma_N,
\end{aligned}
\end{cases}
\end{align}

where $P(u)=\frac{\partial W(F(u))}{\partial F}$ is the first Piola-Kirchhoff stress tensor and $F=I+\nabla u$ the deformation gradient. Following the formulation in \cite{Mendizabal2020-em}, we assume $f = 0$ in the current setting since no volumetric body forces are applied and the deformation is driven solely by boundary conditions and external surface tractions. $W$ represents the strain energy density associated with the chosen \textbf{hyperelastic material model}, and characterizes the nonlinear constitutive behavior of the material.

\section{\textsc{HyperShape}}
\label{sec:hypershape}

This section presents the data generation pipeline proposed in \textsc{\textbf{HyperShape}} (overview in Figure~\ref{fig:hypershape}). We describe the general procedure for sampling shapes and boundary conditions and for constructing the corresponding hyperelastic simulation problems. While the proposed framework is general and not restricted to a specific constitutive law, in this work, we consider a Neo-Hookean \cite{Rivlin1948} material model for all experiments. Given the sampled shapes, we compute solutions to the hyperelasticity problem~\eqref{eq:bvp} using the open-source finite element software FEniCS \cite{baratta_dolfinx_2023, scroggs_construction_2021, scroggs_basix_2022, alnaes_unified_2012}.


\subsection{Shape sampling}\label{sect:shape-sampling}
\begin{figure}[!h]
    \centering

    \begin{minipage}{0.68\linewidth}
        \centering
        \includegraphics[width=\linewidth]{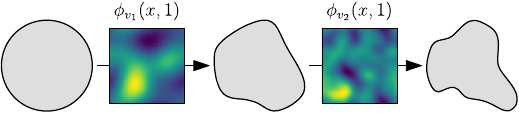}
        \caption{Example of the random shape generation process. Starting from a circle, we apply \textbf{$K=2$} deformations: a coarse followed by a higher-frequency deformation to introduce local details.}
        \label{fig:random_deformations}
    \end{minipage}
    \hfill
    \begin{minipage}{0.30\linewidth}
        \centering
        \includegraphics[width=0.65\linewidth]{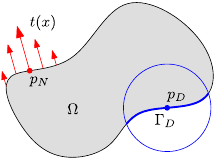}
        \caption{Illustration of the boundary condition sampling procedure.}
        \label{fig:domain_example}
    \end{minipage}

\end{figure}

The basic approach for sampling a shape is that we start with an initial shape that is always the same, and then apply a random deformation to it. In most cases, we use a circle or a sphere as an initial shape. However, this choice is not restrictive, and the initial geometry can be informed by prior knowledge of the target shape distribution. We first describe a method for applying a random deformation to a shape, which can be split into two steps. In the first step, we sample a stationary velocity field $v\sim\mathcal{G}(0,\mathcal{C})$ from a Gaussian random field with covariance:
\begin{equation}
    \mathcal{C}(v(x),v(y))=m^2e^{-\|x-y\|_2^2/(2\sigma^2)}
    \label{eq:cov}
\end{equation}
for given parameters $m,\sigma$. Then, in the second step, we compute the deformation $\tilde{x}=\phi(x,1)$, where $\phi:\mathbb{R}^d\times[0,1]\to\mathbb{R}^d$ is the solution to the following ODE:

\begin{equation}
    \frac{d\phi(x,t)}{dt}=v(\phi(x,t)),\quad\phi(x,0)=x.
\end{equation}
This can be solved using the scaling and squaring methods proposed in \cite{arsigny_log-euclidean_2006}. In Equation~\ref{eq:cov}, the parameter $m$ controls the magnitude of the deformation, while $\sigma$ describes the scale. This means that for lower $\sigma$, the velocity field has a lot of variation over a smaller distance, while for larger $\sigma$, the velocity field changes more slowly, leading to more global deformations. We apply this deformation process $K$ times, with $K \in \{1,2,3\}$, to progressively build shape complexity. In practice, higher-frequency deformations (small $\sigma$) are introduced in later steps to refine local details, while lower-frequency components (large $\sigma$) capture the global structure, as illustrated in Figure~\ref{fig:random_deformations}.


\subsection{Boundary conditions}\label{sect:boundary-conditions}
We consider homogeneous Dirichlet boundary conditions as well as Neumann boundary conditions, i.e., surface traction. We first uniformly sample a point on the boundary of the domain, i.e., the surface of the shape, $p_D\sim U(\partial\Omega)$. \textbf{The Dirichlet boundary} is then defined as the region on the boundary at most distance $r$ away from the point $p_D$, i.e.\ $\Gamma_D:=B_r(p_D)\cap\partial\Omega$. \textbf{The Neumann boundary} is simply $\Gamma_N=\partial\Omega\setminus\Gamma_D$. To define the traction $t:\Gamma_N\to\mathbb{R}^d$, we uniformly sample a point on the surface $p_N\sim U(\partial\Omega\setminus\Gamma_D)$ and set $t(x)=\alpha\cdot\vec{g}\cdot\frac{\varphi(x-p_N)}{\int_{\Gamma_N}\varphi(y)dy}$, where $\alpha\in\mathbb{R}_{\geq0}$ is a scaling factor, $\vec{g}$ is a random vector sampled from $\mathcal{N}(0,I)$ and normalized to 1, and $\varphi(x)=\exp(-\|x\|_2^2/(2\gamma^2))$ distributes the traction over a small region around $p_N$, $\gamma$ is a parameter that describes the concentration of the traction around $p_N$. The scaling $\alpha$ can either be given as a parameter or adjusted to achieve a given displacement magnitude. See Figure~\ref{fig:domain_example}.

\subsection{Data representation}\label{sect:data-representation}

There are different ways neural operators encode their input function. In this study, we focus on two of the most common ways: grid samples and point clouds.

\textbf{Grid-based representation:}
We represent the input shape as a signed distance field (SDF) evaluated on the grid over the domain $[-1,1]^d$. This is possible since all shapes are contained in $[-1,1]^d$. Similarly, the Dirichlet boundary $\Gamma_D$ is also encoded as a signed distance field. Since $\Gamma_D$ is not a closed surface, its SDF has no negative values, just $0$ for points on the surface. Lastly, we need to encode the Neumann boundary condition. Since $\Gamma_N=\partial\Omega\setminus\Gamma_D$, the information on the boundary itself is already given. The domain $\Gamma_N$ of the traction force is extended to $[-1,1]$ and $t$ is given as a field $t:[-1,1]^d\to\mathbb{R}^d$ that is evaluated on the same grid as the other functions. The output is the displacement field, also extended to the domain $[-1,1]^d$ and evaluated on the same grid. To evaluate the prediction, we interpolate the function values at the points of the original mesh.

\textbf{Point cloud representation:}
Our simulations were performed on volumetric meshes, therefore, we can directly use the nodes of the mesh as an input point cloud. However, it is difficult to determine which points lie on the surface of the shape solely from their spatial coordinates. To address this, we augment the point cloud with signed distance function (SDF) values evaluated at each point, enabling the identification of surface regions. In addition, we encode the Dirichlet boundary by including its corresponding SDF values in the same manner. Finally, to ensure consistency with the grid-based representation, the traction field $t$ is extended to all points in the point cloud, including interior points, where it is set to zero outside the Neumann boundary.

\subsection{Datasets generation}

We used \textbf{\textsc{HyperShape}} to generate a suite of 2D and 3D datasets (see Table~\ref{tab:dataset-overview}). In general, we set a single Dirichlet boundary condition for each shape in the dataset and then perform a specified number of simulations with randomly selected surface tractions, as described in Section~\ref{sect:boundary-conditions}. For all simulations, we set the scaling of the traction $\alpha$ such that the simulations for each shape cover a uniform range of displacement magnitudes from $0.1$ to $0.5$. Since all shapes are normalized to the domain $[-1,1]^d$, the largest displacement magnitude is therefore $1/4$ of the domain's length.

\begin{figure}[!h]
    \centering
    \begin{minipage}{0.48\linewidth}
        \centering
        \includegraphics[width=1.\linewidth]{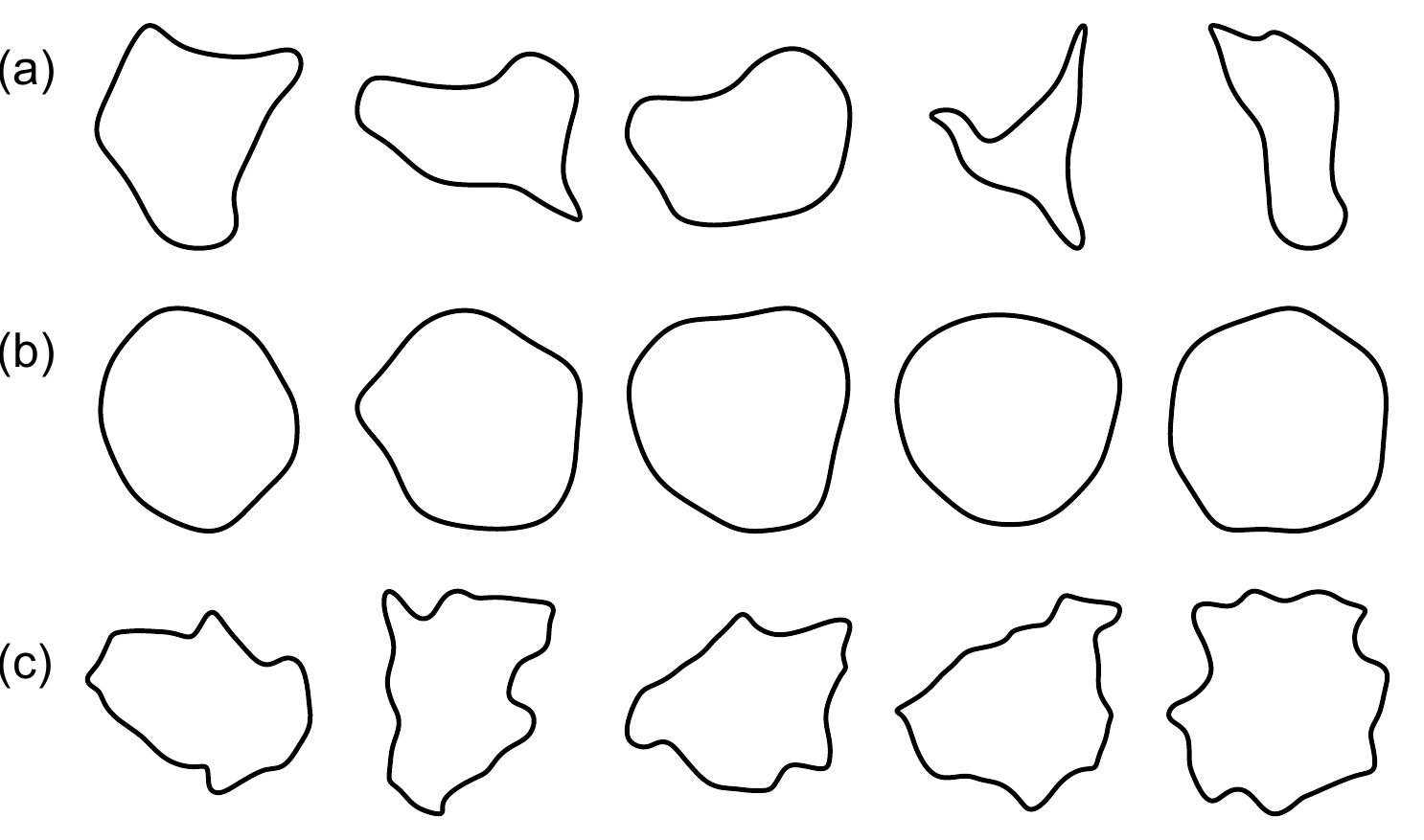}
        \caption{Examples of generated 2D shapes from \textbf{Random2D} (a), \textbf{EasyRandom2D} (b) and (c) \textbf{OODRandom2D}.}
        \label{fig:shapes2d}
    \end{minipage}
    \hfill
    \begin{minipage}{0.48\linewidth}
        \centering
        \includegraphics[width=1.\linewidth]{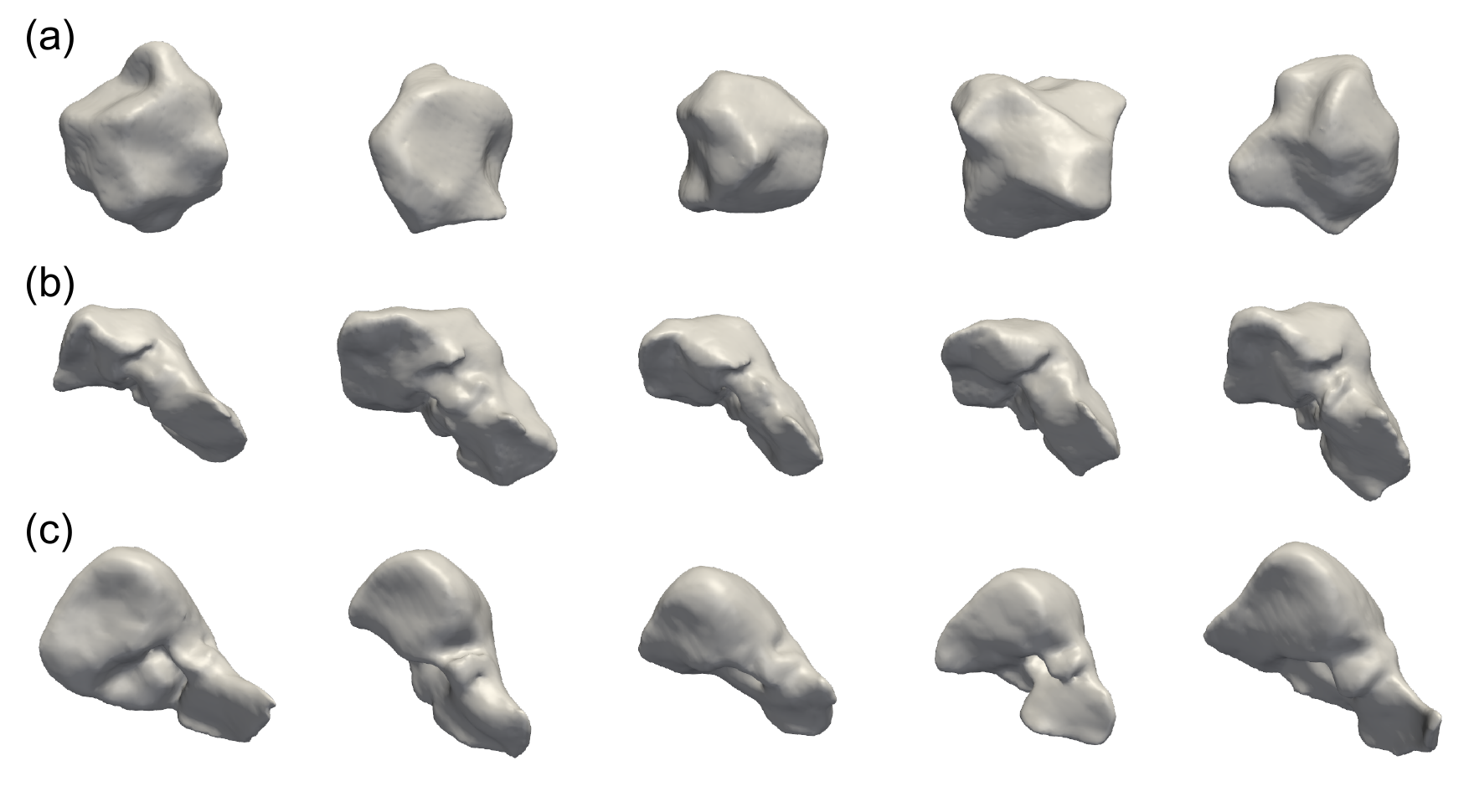}
        \caption{Examples of generated 3D shapes from (a) \textbf{Random3D}, (b) \textbf{AugLiver} and (c) \textbf{Liver}.}
        \label{fig:shapes3d}
    \end{minipage}
\end{figure}

For the 2D datasets, \textbf{\textsc{HyperShape}} follows the procedure described in Section~\ref{sect:shape-sampling}, where shapes are generated by applying random deformations to an initial circular geometry. Using this pipeline, we generate four datasets (\textbf{Random2D}, \textbf{EasyRandom2D}, \textbf{OODRandom2D}, and \textbf{OneRandom2D} see Table~\ref{tab:dataset-overview}), which all share the same circular base shape but differ in the number of distinct shapes, the number of simulations per shape, and the deformation parameters controlling complexity and distribution. The specific parameters used for shape generation are reported in the last column of Table~\ref{tab:dataset-overview}, and examples are shown in Figure~\ref{fig:shapes2d}. For consistency, the same boundary condition parameters are used across all 2D datasets ($r=0.5$ and $\gamma=0.2$), with a grid resolution of 128.

For the 3D datasets, we consider both purely synthetic shapes and anatomically derived geometries. Specifically, we use $92$ liver meshes obtained from segmentations of the publicly available Liver Tumor Segmentation Dataset \cite{bilic_liver_2023} (CC BY-NC-ND 4.0). Based on these, we construct two datasets: one containing all available liver shapes (\textbf{Liver}), and another restricted to a single liver geometry (\textbf{OneLiver}). In addition, we used \textbf{\textsc{HyperShape}} to generate synthetic 3D datasets. This includes (i) generic shapes obtained by applying \textbf{\textsc{HyperShape}} with a sphere as a base shape (\textbf{Random3D}), and (ii) a synthetic liver dataset generated by using \textbf{\textsc{HyperShape}} with a real liver as a base shape (\textbf{AugLiver}). Example shapes are shown in Figure~\ref{fig:shapes3d}. For consistency across experiments, we used the same boundary condition parameters for all 3D datasets ($r=0.4$, $\gamma=0.1$),  and grid resolution of $64$. 


\begin{table}
  \centering
  \caption{Overview of the datasets generated with \textbf{\textsc{HyperShape}}, including the number of shapes, simulations per shape, base geometry, and deformation parameters. They vary in geometric complexity and are designed to evaluate in-distribution, out-of-distribution, and single-shape settings.}
  \label{tab:dataset-overview}
  \begin{tabular}{l l c c c c}
    \toprule
    Dim & Name & \# Shapes & \# Sim./Shape & Base & Params $\{(\sigma,m)\}$ \\
    \midrule
    \multirow{4}{*}{2D} 
      & Random2D     & 1000 & 50     & Circle  & $(0.2,4),(0.1,1)$ \\
      & EasyRandom2D & 500  & 50     & Circle  & $(0.1,0.5)$ \\
      & OODRandom2D  & 500  & 50     & Circle  & $(0.2,2),(0.05,1)$ \\
      & OneRandom2D  & 1    & 10\,000 & Circle  & $(0.2,4),(0.1,1)$ \\
    \midrule
    \multirow{4}{*}{3D} 
      & Random3D     & 1000 & 20     & Sphere  & $(0.2,4),(0.1,4),(0.05,1)$ \\
      & AugLiver     & 500  & 20     & Liver   & $(0.2,2),(0.1,1),(0.05,1)$ \\
      & Liver        & 92   & 100    & --      & -- \\
      & OneLiver     & 1    & 10\,000 & --      & -- \\
    \bottomrule
  \end{tabular}
\end{table}

\section{Experiments}
\label{sec:experiments}

In this section, we describe several experiments that investigate different aspects of generalizing hyperelasticity to distributions of shapes. For the experiments, we consider the grid-based neural operators FNO \cite{li_fourier_2021,kossaifi_multi-grid_2023}, CNO \cite{raonic_convolutional_2023} and a modified version of the U-Net \cite{ronneberger_u-net_2015, liu_convnet_2022} (described in Appendix~\ref{app:u-net}), and point-cloud-based neural operators GINO \cite{li_geometry-informed_2023} and Transolver \cite{wu_transolver_2024}. Each model has 2D and 3D versions that remain the same across all experiments and whose hyperparameters were primarily taken from the standard settings of the respective papers introducing them (see Appendix~\ref{app:hyperparams}). Unless stated otherwise, we use an 80-10-10 training-validation-test split of the datasets, with each shape appearing in only one split. The loss function and the evaluation metrics were computed from the predicted displacement values $\hat{u}=\{\hat{u}^i\}_{i=1}^N$ at the mesh points used in the FEM simulations, with corresponding ground-truth values $u=\{u^i\}_{i=1}^N$. Following previous work \cite{wu_transolver_2024, boulle_chapter_2024, duprez_phi-fem-fno_2025}, we define a discretized version of $L^2$ and $H^1$ norm respectively as:

\begin{equation}
\|u\|_{L^2_N}
:=\sqrt{\frac{1}{N_p}\sum_{i=1}^{N_p}\|u^i\|_2^2},
\quad
\|u\|_{H^1_N}
=\sqrt{\frac{1}{N_p}\sum_{i=1}^{N_p}\left(\|u^i\|_2^2
+\sum_{k=1}^d\|\nabla_{x_k}u^i\|_2^2\right)}
\end{equation}

We then can define standard error metrics such as Root Mean square error as  $\mathcal{L}_{L^2}(u,\hat{u}):=\|u-\hat{u}\|_{L^2_N}$, also called $L^2$-loss, and its analogue including derivatives $\mathcal{L}_{H^1}(u,\hat{u}):=\|u-\hat{u}\|_{H^1_N}$, which we simply call $H_1$-loss. The corresponding relative errors are defined as $\mathcal{L}_{L^2}^{\mathrm{rel}}(u,\hat{u}):=\|u-\hat{u}\|_{L^2_N}/\|u\|_{L^2_N}$ and $\mathcal{L}_{H^1}^{\mathrm{rel}}(u,\hat{u}):=\|u-\hat{u}\|_{H^1_N}/\|u\|_{H^1_N}$. For training, grid-based models are optimized with the relative $H^1$-loss $\mathcal{L}_{H^1}^{\mathrm{rel}}$, since spatial derivatives can be approximated using finite differences. In contrast, point-based models are trained using the relative $L^2$-loss $\mathcal{L}_{L^2}^{\mathrm{rel}}$.

To assess the correlation between geometric complexity and prediction error, we use two measures. \textbf{In 2D}, we use the isoperimetric ratio \cite{osserman_isoperimetric_1978} to characterize the complexity of shapes. Given a 2D shape with a closed curve of length $P$ and an area $A$. The \textbf{isoperimetric ratio} equals $4\pi A/P^2$. It measures how close a given shape is to a circle. The circle maximizes this ratio with $1$. Meanwhile, by considering a star shape and increasing the number of spikes in the star, we can get arbitrarily close to $0$. \textbf{In 3D}, we use the Hausdorff distance. Since this metric is generally scale-dependent, all shapes are normalized to the domain $[-1,1]^d$ to ensure comparable distances across geometries.



\subsection{In-distribution benchmark}\label{sect:in-dist}

\begin{figure}[!h]
    \centering
    \includegraphics[width=\linewidth]{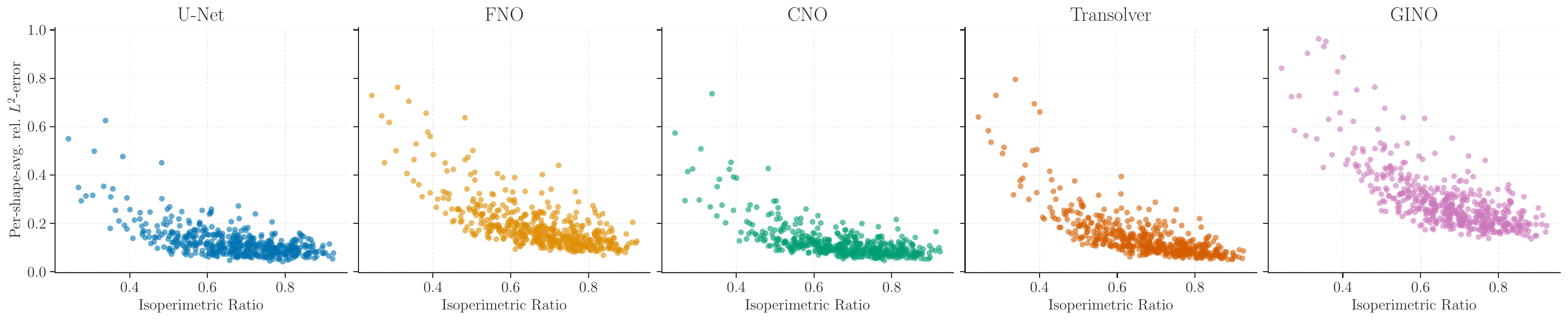}
    \caption{Relationship between geometric complexity and prediction error. Each plot shows the isoperimetric ratio (x-axis) against the per-shape average relative $L^2$-error on the test split (500 shapes) of the \textbf{Random2D} dataset, for five different models. Across all cases, shapes closer to a circular geometry (isoperimetric ratio near 1) consistently yield lower prediction errors.}
    \label{fig:isoperimetric_vs_error}
\end{figure}

In this section, we evaluate the ability of the models to generalize within a given data distribution when trained and tested on samples drawn from the same \textbf{\textsc{HyperShape}} setting. As mentioned previously, between the 2D and 3D datasets, only the distribution of geometries is different, while the sampling process for the boundary conditions is the same. This allows us to isolate and study the effect of increasing geometric complexity. As shown in Table~\ref{tab:in-dist}, all models exhibit a clear degradation in performance as the variability of the underlying geometry increases, moving from the single-shape setting (\textbf{OneRandom2D} / \textbf{OneLiver}), to slightly varying geometries (\textbf{EasyRandom2D} / \textbf{Liver}), and finally to more diverse shape distributions (\textbf{Random2D} / \textbf{Random3D}). This highlights the increasing difficulty of the problem as geometric variability grows. In the 2D case, point-cloud-based methods achieve the best performance in the single-shape setting, whereas grid-based methods become more competitive when multiple shapes are introduced. A similar behavior is observed in 3D, where all models perform best in the fixed-geometry setting and show increased errors as the number and variability of shapes grow. To further analyze this effect, Figure~\ref{fig:isoperimetric_vs_error} shows the relationship between the isoperimetric ratio and prediction error on the \textbf{Random2D} test set. Across all models, more regular shapes (higher isoperimetric ratio, closer to a circle) consistently yield lower errors, while more complex shapes lead to higher errors, confirming the strong impact of geometry on performance.



\begin{table}[!h]
  \centering
  \caption{In-distribution performance on \textsc{HyperShape} datasets, reported as mean relative $L^2$-error. We compare models across increasing geometric variability, from single-shape to diverse shape settings. Best results are in \textbf{bold} and second-best are \underline{underlined}}
  \label{tab:in-dist}
  \begin{tabular}{ccccccc}
    \specialrule{1.2pt}{0pt}{0pt}

    \multirow{2}{*}{Dim} & \multirow{2}{*}{Dataset}
    & \multicolumn{3}{c}{Grid}
    & \multicolumn{2}{c}{Point-cloud} \\
    
    \cmidrule(lr){3-5} \cmidrule(lr){6-7}

    & 
    & U-Net \cite{ronneberger_u-net_2015} 
    & FNO \cite{li_fourier_2021} 
    & CNO \cite{raonic_convolutional_2023} 
    & GINO \cite{wen_geometry_2026} 
    & Transolver \cite{wu_transolver_2024} \\

    \midrule

    \multirow{3}{*}{2D} 
      & OneRandom2D  & 0.0183 & 0.0191 & 0.0146 & \underline{0.0087} & \textbf{0.0057} \\
      & EasyRandom2D & \textbf{0.0293} & 0.0599 & 0.0588 & 0.0965 & \underline{0.0391} \\
      & Random2D     & \underline{0.1209} & 0.1886 & \textbf{0.1180} & 0.2921 & 0.1429 \\

    \midrule

    \multirow{3}{*}{3D} 
      & OneLiver     & \underline{0.0402} & 0.0474 & 0.0424 & 0.0440 & \textbf{0.0378} \\
      & Liver        & 0.1915 & \underline{0.1843} & \textbf{0.1677} & 0.2179 & 0.1886 \\
      & Random3D     & \underline{0.1295} & 0.1498 & \textbf{0.1271} & 0.1766 & 0.1484 \\

    \specialrule{1.2pt}{0pt}{0pt}
  \end{tabular}
\end{table}

\subsection{Out-of-distribution benchmark}

In this part, we investigate the ability of neural operators to generalize to shapes that are out of the training distribution but still arguably close enough. We address this question using a specific use case. We perform experiments for 2D and 3D cases and the results are reported in Table~\ref{tab:ood}.

\paragraph{2D case:} We first evaluated out-of-distribution generalization in 2D. Models were trained on subsets of the \textbf{Random2D} and \textbf{EasyRandom2D} datasets (see Section~\ref{sect:in-dist}) and then tested on the \textbf{OODRandom2D} dataset, which consists of shapes drawn from a different \textsc{\textbf{HyperShape}} generation setting (see Table~\ref{tab:dataset-overview}) and therefore from a different distribution. This setup allowed us to measure how well models generalize beyond the geometric variations seen during training. As shown in Table~\ref{tab:ood}, in the 2D setting, we observe a clear improvement in out-of-distribution performance when models are trained on the more diverse \textbf{Random2D} dataset compared to the simpler settings.

\paragraph{3D case:} We then consider the 3D setting, where we define three training scenarios. First, models are trained on fully synthetic shapes from \textbf{Random3D}. Second, they are trained on \textbf{AugLiver}, which consists of a single liver shape augmented with random deformations, thereby more closely reflecting anatomical variability. Finally, models were trained directly on liver shapes. We evaluated models trained on \textbf{Random3D} on the full \textbf{Liver} dataset, and models trained on \textbf{AugLiver} on all liver shapes except the original base shape used for augmentation. The results for models trained directly on $80\%$ of the \textbf{Liver} and tested on $10\%$ of the same dataset are reported, serving as a reference. These experiments aimed at assessing how well synthetic pretraining transfers to real anatomical geometries. As observed in Table~\ref{tab:ood}, point-cloud-based models struggle significantly when evaluated on shapes that differ substantially from the training distribution. In particular, models trained on \textbf{Random3D} fail to produce meaningful predictions on the \textbf{Liver} dataset, while performance improved substantially when trained on \textbf{AugLiver}, whose shapes are closer to the target anatomical distribution. In contrast, grid-based models such as \textbf{U-Net}, \textbf{FNO}, and \textbf{CNO} performed better when trained on \textbf{Random3D}. One possible explanation is that \textbf{Random3D} contains a larger number of training samples than \textbf{AugLiver}, providing better coverage of shape variability.

\begin{figure}
    \centering
    \includegraphics[width=\linewidth]{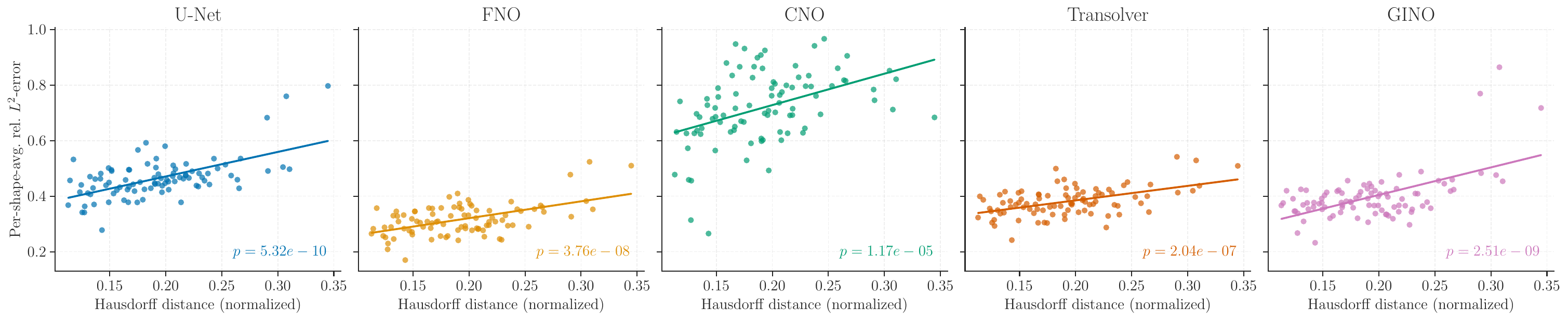}
    \caption{Shape-averaged relative $L^2$-error for the remaining 91 liver shapes from the Liver dataset plotted against the normalized Hausdorff distance of the respective shapes to the first liver shape. Evaluated models were trained on AugLiver, which uses the first liver shape as the base shape.}
    \label{fig:hausdorff_vs_error}
\end{figure}

To analyze the correlation between the geometric variability and prediction accuracy, since the $92$ shapes in the \textbf{Liver} dataset are registered, we compute the Hausdorff distance between the liver shape used as the base in \textbf{AugLiver} and each of the remaining 91 shapes. We then compared this geometric discrepancy with the corresponding prediction errors, as shown in Figure~\ref{fig:hausdorff_vs_error}. The results show a clear trend: as the Hausdorff distance increases, the relative $L^2$ error also increases consistently across all models. This indicates that model accuracy gradually deteriorates as the target shape deviates further from the training geometry.

\begin{table}
  \centering
  \caption{Model comparison under out-of-distribution and transfer settings. Models trained on \textbf{Random2D} and \textbf{EasyRandom2D} are evaluated on \textbf{OODRandom2D}. For the 3D transfer setting, models are trained on \textbf{Random3D}, \textbf{AugLiver}, and \textbf{Liver}, and evaluated on the Liver dataset. Reported is the mean relative $L^2$-error. Best results are in \textbf{bold} and second-best are \underline{underlined}.}
  \label{tab:ood}
  \begin{tabular}{llccccc}
    \toprule

    Dim & Dataset 
    & \multicolumn{3}{c}{Grid}
    & \multicolumn{2}{c}{Point-cloud} \\

    \cmidrule(lr){3-5} \cmidrule(lr){6-7}

    & 
    & U-Net \cite{ronneberger_u-net_2015}
    & FNO \cite{li_fourier_2021}
    & CNO \cite{raonic_convolutional_2023}
    & GINO \cite{wen_geometry_2026}
    & Transolver \cite{wu_transolver_2024} \\

    \midrule

    \multirow{2}{*}{2D}
      & Random2D     & \textbf{0.0991} & 0.1551 & 0.1082 & 0.2467 & \underline{0.1160} \\
      & EasyRandom2D & \textbf{0.1893} & 0.3393 & 0.2418 & 0.5240 & \underline{0.2167} \\

    \midrule

    \multirow{3}{*}{3D}
      & Random3D     & \textbf{0.3347} & \underline{0.3621} & 0.3654 & 1.4073 & 0.8096 \\
      & AugLiver     & 0.4641 & \textbf{0.3160} & 0.7191 & 0.3967 & \underline{0.3807} \\
      & Liver        & 0.1915 & 0.1843 & \textbf{0.1677} & 0.2179 & \underline{0.1886} \\

    \bottomrule
  \end{tabular}
\end{table}

\subsection{Impact of training data size}

\begin{figure}[!h]
    \centering
    \includegraphics[width=0.95\linewidth]{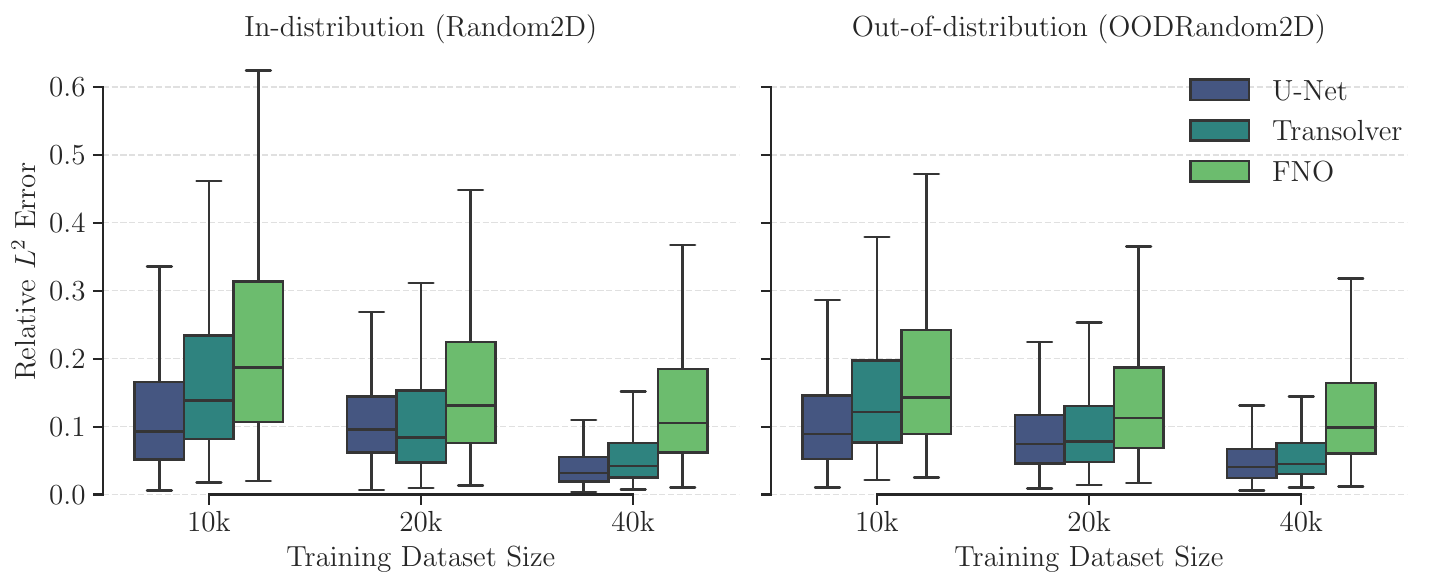}
    \label{fig:sample-efficiency}
    \caption{Data efficiency of \textbf{U-Net}, \textbf{FNO}, and \textbf{Transolver} trained on increasing dataset sizes and evaluated on in-distribution (\textbf{Random2D}) and out-of-distribution (\textbf{OODRandom2D}) settings. Results show consistent gains with more data for both distributions.}
\end{figure}

A key question is how much training data is required for models to generalize over a given distribution of problems. We focus on the \textbf{U-Net}, \textbf{FNO}, and \textbf{Transolver} models, and evaluate their performance when trained on increasing amounts of data (10K, 20K, and 40K samples), corresponding to $200$, $400$, and $800$ shapes respectively. We assess data efficiency by measuring test error on a hold-out set from the same distribution and on the out-of-distribution dataset \textbf{OODRandom2D}, thereby evaluating how scaling the training data affects both in- and out-of-distribution performance. As shown in Figure~\ref{fig:sample-efficiency}, all models consistently improve as the amount of training data increases, with gains still visible beyond 20,000 samples for both in-distribution and out-of-distribution test sets. This indicates that current models require large training datasets to fully capture geometric variability and achieve strong generalization.

\section{Conclusion and limitations}
\label{sec:conclusion}

We introduced \textbf{\textsc{HyperShape}}, a controllable and extensible framework for generating synthetic hyperelastic datasets in 2D and 3D, designed for in-distribution, out-of-distribution, and synthetic-to-real evaluation. In contrast to existing benchmarks, \textbf{\textsc{HyperShape}} enables controlled variation in both shape complexity and physical setup, including boundary conditions and applied forces, providing a more realistic testbed for hyperelastic surrogate models. 
We used \textbf{\textsc{HyperShape}} to generate multiple 2D and 3D datasets with increasing geometric complexity, ranging from single-shape settings to diverse synthetic and anatomical shape distributions, and used them to benchmark five state-of-the-art neural operators. The results show a consistent performance degradation as the geometric diversity increases. Out-of-distribution and synthetic-to-real experiments further highlight the limited robustness of current models, which remain highly sensitive to the training distribution and require large amounts of data to perform reliably. These findings emphasize the need for further development of geometry- and physics-aware models and position \textbf{\textsc{HyperShape}} as a benchmark for hyperelastic surrogate modeling. Yet, we envision several areas of improvement for our framework. In its present form, \textbf{\textsc{HyperShape}} is restricted to homogeneous materials, which excludes important applications such as heterogeneous soft tissue simulation. Additionally, the influence of material parameters and their variability on model performance remains unexplored. Another limitation concerns boundary conditions and applied forces, which are currently constrained to a single surface location; enabling multiple simultaneous forces at different locations would further improve realism. Finally, while we observe that increasing geometric complexity generally leads to reduced model performance, the joint effect of geometry and boundary conditions is not yet fully understood. In particular, these factors do not act independently: different levels of geometric complexity can amplify or mitigate the difficulty introduced by specific boundary conditions or force configurations. As a result, their combined influence leads to varying degrees of problem difficulty, making it challenging to isolate which aspect is primarily responsible for performance degradation.


\bibliographystyle{abbrvnat}
\bibliography{ref}


\appendix

\section{Simulation details}

\subsection{Material model}
Hyperelastic material can be modeled in different ways. In this work, we consider the commonly used Neo-Hookean model \cite{grundemann_ogden_1985}, which defines the strain energy density function as
\[
    W=\frac{\mu}{2}(\tr(F^\top F)-3-2\ln(\det F))+\frac{\lambda}{2}(\ln(\det F))^2.
\]
The scalars $\mu,\lambda$ are so-called Lamé parameters and can be expressed by $\mu=E/(2(1+\nu)),\lambda=E\nu/((1+\nu)(1-2\nu))$, where $E$ is the Young's modulus and $\nu$ the Poisson ratio. Both $E$ and $\nu$ are material parameters and can vary between different materials. The Poisson ratio $\nu$ is between $0$ and $0.5$. The closer $\nu$ is to $0.5$ the more volume preserving it is. For our datasets, we keep the parameters fixed at $E=500~(\mathrm{Pa})$ and $\nu=0.49$. These parameters can be easily changed in the configuration file for the data generation. Furthermore, also the material model can be replaced by changing a few lines of code and writing a different formula for $W$.

\subsection{Dataset size}
We already report the size of the dataset in Table\ref{tab:dataset-overview}. However, in a few cases, the FEM simulations are unable to converge. This means that some of the shapes in \textbf{Random2D} have a few simulations less than reported in the table. Concretely the total number of simulations in \textbf{Random2D} is 49'924. We believe that this does not have a meaningful impact on our results and conclusions. In all other datasets the number of simulations is complete and as reported.

\subsection{Liver shape extraction}
The liver geometries used in this work were derived from the binary segmentation masks provided in the Liver Tumor Segmentation Dataset \cite{bilic_liver_2023}. For each segmentation, a thresholding operation was first applied to isolate the liver region. Surface meshes were then extracted from the resulting binary volumes using the Marching Cubes algorithm \cite{Lorensen1987-jx}, which produces a triangulated surface mesh from the iso-surface of the segmentation mask. To remove the staircase artifacts inherent to voxel-based segmentations, Laplacian smoothing \cite{Field1988-fy} was subsequently applied to the extracted meshes. Finally, Poisson surface reconstruction \cite{kazhdan_poisson_2006} was performed to obtain a clean, watertight surface mesh suitable for FEM simulation. The resulting meshes were then registered using iterative closest point \cite{Besl1992-zt}.

\section{Experiment details}

\subsection{Training}
For every training, the models are validated every 10 epochs of the training on a hold-out validation split (always 10\% of the dataset). Over the whole training run, the checkpoint performing best on the validation split is selected for testing. For the training on \textbf{Random2D} as well as the data efficiency study on \textbf{Random2D}, the number of training epochs is 500. For all other trainings, the number of epochs is 200. All models use the AdamW optimization algorithm \cite{loshchilov2019decoupledweightdecayregularization}. The specific scheduler as well as the other hyperparameters depend on the model and are reported in Appendix~\ref{app:hyperparams}.
The training dataset split is 80\% of the dataset, unless stated otherwise. For the training of \textbf{Random2D}, a 40\% training split is used, resulting in 400 shapes, which is the same number of shapes as used in the training on \textbf{EasyRandom2D}.

\subsection{Comparison to Elasticity benchmark}
Here we explain the results in Figure~\ref{fig:elasticity-benchmark-comparison} depicting the comparison between our Random2D dataset and the popular hyperelasticity benchmark \cite{li_fourier_2024}, which we call Elasticity. The neural operator models have the same hyperparameter configurations as elsewhere. While the original benchmark only provides the point cloud of the geometry itself, we took a similar approach as described in Section~\ref{sect:data-representation}, where we add the signed distance function values of the given geometry to the input. This means that the point-cloud-based models are given the point cloud and the SDF value of the shape, while the grid-based model are simply given the SDF evaluated on the grid.

\subsection{Computational resources}\label{app:computational-resources}
All deep learning experiments were conducted on NVIDIA A100 (40 GB) and RTX 6000 (24 GB) GPUs. Individual training runs ranged from approximately 0.5 to 5 days depending on the configuration. The experiments reported in the paper required a total of approximately 75-100 GPU-days. FEM-based synthetic data generation was performed on a standard laptop CPU, requiring approximately 12 to 48 hours per dataset, depending on its size. Beyond the reported experiments, additional training runs were conducted during the research process; the total compute including these is difficult to estimate precisely but was likely higher than the reported experiments alone.

\section{Model details}

\subsection{Modified U-Net}\label{app:u-net}
Our U-Net model is mainly based on the standard U-Net architecture \cite{ronneberger_u-net_2015}, but with two important modifications. First, we replace standard convolution blocks by ConvNeXt convolutions \cite{liu_convnet_2022}. Second, we combine the outputs predicted from all upsampling layers of the U-Net, similar to \cite{lin_feature_2017}. Concretely, we extract multi-scale feature maps from the upsampling blocks of our U-Net, project each to the output channel dimension via 1x1-convolutions, upsample them to the final output resolution, and sum them up to produce the final output. Along with this second modification, we use bilinear interpolation for up- and downsampling throughout the architecture. The intuition for this modification is that we want the model to predict the lower frequency modes of the displacement field in the lower resolutions and adding the details of the displacement fields in the higher resolution layers.

Furthermore, we also use a different normalization method for the U-Net model, where we only normalize the force field and displacement field by the root mean square (RMS) averaged over the complete training dataset. The signed distance functions encoding the shape and Dirichlet boundary are not normalized. We argue that this is a more natural way to normalize the data. For all other models, we use the standard method or channelwise normalization \cite{li_fourier_2021}.

We study the impact of these three modifications as well as the influence of using relative $L^2$-error vs.\ $H^1$-error in a small ablation study. We train U-Net with different settings on the first 400 shapes of the Random2d dataset, use 100 shapes as validation and the remaining 500 for testing. All versions have around 500'000 trainable parameters. The results are in Table~\ref{tab:unet-ablation} and report the average result over three training runs with different seeds.

\begin{table}[ht]
  \caption{Ablation study for the U-Net architecture. The last row represents the final configuration used in all the experiments.}
  \label{tab:unet-ablation}
  \centering
  \begin{tabular}{ccccc}
    \toprule
    ConvNeXt & Multi-scale comb. & Norm & Loss & rel.\ $L^2$-error ($\pm$std)\\
    \midrule
    \ding{55} & \ding{55} & RMS         & rel.\ $H^1$-loss & $0.4534\pm0.0043$\\
    \ding{51} & \ding{55} & RMS         & rel.\ $H^1$-loss & $0.1044\pm0.0042$\\
    \ding{51} & \ding{51} & channelwise & rel.\ $H^1$-loss & \underline{$0.0918\pm0.0035$}\\
    \ding{51} & \ding{51} & RMS         & rel.\ $L^2$-loss & $0.1090\pm0.0018$\\
    \ding{51} & \ding{51} & RMS         & rel.\ $H^1$-loss & $\mathbf{0.0905\pm0.0017}$\\
    \bottomrule
  \end{tabular}
\end{table}

We can observe that the normalization method does not seem to have an influence on performance. However, using the relative $H^1$-loss instead of the $L^2$-loss leads to an improvement. The ConvNeXt modification seems to have the most significant impact, but also the multi-scale feature combination improves the result.

\subsection{Hyperparameter settings}\label{app:hyperparams}

\begin{table}[ht]
  \caption{Hyperparameters for the U-Net \cite{ronneberger_u-net_2015, liu_convnet_2022}.}
  \label{tab:unet-hyperparams}
  \centering
  \begin{tabular}{lcc}
    \toprule
    Hyperparameter & U-Net (2D) & (3D) \\
    \midrule
    Resolution & 128 & 64 \\
    Input / Output Channels & 4 / 2 & 5 / 3 \\
    Channels per Level & [16, 32, 64, 128] & --- \\
    Blocks per Level & [4, 4, 4, 4] & --- \\
    Kernel Size & 7 & --- \\
    Normalization Strategy & RMS & --- \\
    Loss Function & rel-H1 & --- \\
    Batch Size & 16 & --- \\
    Learning Rate & $10^{-3}$ & --- \\
    Weight Decay & $10^{-5}$ & --- \\
    LR Scheduler & cosine & --- \\
    Min. Learning Rate & $10^{-6}$ & --- \\
    \bottomrule
  \end{tabular}
\end{table}

\begin{table}[ht]
  \caption{Hyperparameters for the Transolver model \cite{wu_transolver_2024}.}
  \label{tab:transolver-hyperparams}
  \centering
  \begin{tabular}{lcc}
    \toprule
    Hyperparameter & Transolver (2D) & (3D) \\
    \midrule
    Resolution & 128 & 64 \\
    In / Out / Space Dim. & 4 / 2 / 2 & 5 / 3 / 3 \\
    Number of Layers & 8 & --- \\
    Hidden Dimension & 256 & --- \\
    Slice Number & 64 & --- \\
    Attention Heads & 8 & --- \\
    MLP Ratio & 4 & --- \\
    Normalization Strategy & channelwise & --- \\
    Loss Function & rel-L2 & --- \\
    Batch Size & 16 & 8 \\
    Learning Rate & $10^{-3}$ & --- \\
    Weight Decay & $10^{-5}$ & --- \\
    Grad. Clipping & True (1.0) & --- \\
    LR Scheduler & cosine\_warmup & --- \\
    \bottomrule
  \end{tabular}
\end{table}

\begin{table}[ht]
  \caption{Hyperparameters for the Fourier Neural Operator (FNO) \cite{li_fourier_2021, kossaifi_multi-grid_2023}.}
  \label{tab:fno-hyperparams}
  \centering
  \begin{tabular}{lcc}
    \toprule
    Hyperparameter & FNO (2D) & (3D) \\
    \midrule
    Resolution & 128 & 64 \\
    Input / Output Channels & 4 / 2 & 5 / 3 \\
    Spectral Modes & [24, 24] & [12, 12, 12] \\
    Hidden Channels & 64 & --- \\
    Number of Layers & 6 & --- \\
    Lifting / Projection Ratio & 2 & --- \\
    Factorization & None & tucker \\
    Rank & N/A & 0.4 \\
    Normalization Strategy & channelwise & --- \\
    Loss Function & rel-H1 & --- \\
    Batch Size & 16 & 8 \\
    Learning Rate & $10^{-3}$ & --- \\
    Weight Decay & 0.0 & --- \\
    LR Scheduler & step\_decay & --- \\
    \bottomrule
  \end{tabular}
\end{table}

\begin{table}[ht]
  \caption{Hyperparameters for the Convolutional Neural Operator (CNO) \cite{raonic_convolutional_2023}. Importantly, we use the \texttt{CNO2d\_simplified} implementation \cite{cno_github} and its straight-forward 3d adaptation.}
  \label{tab:cno-hyperparams}
  \centering
  \begin{tabular}{lcc}
    \toprule
    Hyperparameter & CNO (2D) & (3D) \\
    \midrule
    Resolution & 128 & 64 \\
    Input / Output Dim & 4 / 2 & 5 / 3 \\
    Number of Layers & 4 & --- \\
    Residual Blocks & 4 & --- \\
    Neck Res. Blocks & 6 & --- \\
    Channel Multiplier & 32 & --- \\
    Use Batchnorm & True & --- \\
    Normalization Strategy & channelwise & --- \\
    Loss Function & rel-H1 & --- \\
    Batch Size & 16 & 8 \\
    Learning Rate & $10^{-3}$ & --- \\
    Weight Decay & $10^{-6}$ & --- \\
    LR Scheduler & step\_decay & --- \\
    \bottomrule
  \end{tabular}
\end{table}

\begin{table}[ht]
  \caption{Hyperparameters for the Geometry-Informed Neural Operator \cite{li_geometry-informed_2023}.}
  \label{tab:gino-hyperparams}
  \centering
  \begin{tabular}{lcc}
    \toprule
    Hyperparameter & GINO (2D) & (3D) \\
    \midrule
    Resolution & 128 & 64 \\
    Input / Output Channels & 4 / 2 & 5 / 3 \\
    Latent Grid Resolution & [64, 64] & [32, 32, 32] \\
    GNO Radius & 0.09 & --- \\
    FNO Modes & [24, 24] & [12, 12, 12] \\
    FNO Layers / Hidden & 6 / 64 & --- \\
    FNO Factorization & None & tucker \\
    FNO Rank & N/A & 0.4 \\
    GNO MLP Width (In/Out) & 64 / 64 & --- \\
    Loss Function & rel-L2 & --- \\
    Batch Size & 4 & 8 \\
    Learning Rate & $2 \cdot 10^{-4}$ & --- \\
    Weight Decay & 0.0 & --- \\
    LR Drop Epochs & [50, 100, 200, 300, 400] & [50, 100] \\
    \bottomrule
  \end{tabular}
\end{table}

\clearpage 



\end{document}